# Superconductivity in single crystalline Pb nanowires contacted by normal metal electrodes

Jian Wang[1,2]*, Yi Sun[2]*, Mingliang Tian[1,3], Bangzhi Liu[4], Meenakshi Singh[1], Moses H. W. Chan[1]*

[1]The Center for Nanoscale Science and Department of Physics, The Pennsylvania State University, University Park, Pennsylvania 16802-6300, USA

[2]International Center for Quantum Materials and State Key Laboratory for Mesoscopic Physics, School of Physics, Peking University, Beijing, 100871, People's Republic of China

[3]High Magnetic Field Laboratory, Chinese Academy of Science, Hefei 230031, Anhui, People's Republic of China

[4]Materials Research Institute, The Pennsylvania State University, University Park, Pennsylvania 16802-6300, USA

The transport properties of superconducting single crystal Pb nanowires of 55 nm and 70 nm diameter are studied by standard four electrodes method. Resistance-temperature (*R-T*) scans and magneto-resistance (*R-H*) measurements show a series of resistance steps with increasing temperature and magnetic field as the wires are brought toward the normal state. The resistance-current (*R-I*) scans at different temperature and magnetic field show that the increase in *R* with *I* is punctuated with sharp steps at specific current values. We interpret these steps as consequence of phase slip centers (PSCs) in the superconducting wires enhanced by the presence of the normal Pt electrodes.

## I. INTRODUCTION

The study of superconductivity in nanowires and quasi-one-dimensional (quasi-1D) nanostructures is driven both by open questions in these systems and also the potential applications in dissipationless electronic devices [1-16]. Low dimensional Pb nanostructures have been extensively studied for decades [5-12, 16-18]. Additionally, amorphous and granular nanowires of Pb and other materials have been studied systematically [1, 17-21]. In the last few years, there are a number of experiments studying the properties of single crystal superconducting nanowires with diameter less than 100 nm [13,14,22-24]. An overarching theme of these studies is to understand how superconductivity in these wires goes away with decreasing diameter. It is known that superconductivity is not possible in a true 1D system. A superconducting nanowire approaches the one dimensional (1D) limit when its diameter is reduced towards the Ginzburg-Landau phase coherence length and the magnetic penetration depth [25]. It has been predicted that in approaching the 1D limit superconductivity in the wire is lost near the (bulk) transition temperature via the thermally activated phase slip (TAPS) process [26-28] and at low tempearture via the quantum phase slip (QPS) tunneling process [2,3,29,30]. Experimental evidence of TAPS have been observed in a number of experiments [31-34]. Experimental reports of QPS are not universally accepted [1,4,13,21,34-39]. Local crystalline defects in nanowires [13,23,40-42], whiskers [43,44] and microbridges [45] also play an important role in the sueprconducting property of these systems. When connected to superconducting electrodes, these defects or weak spots act as phase slip centers (PSCs) [42] and give rise to resistance steps in transport measurements with increasing temperature and excitation current.

The transport properties of a superconducting n anowire (and indeed any nanowire) is expected to be strongly influenced by the electrodes contacting the wire. This electrode effect on crystalline nanowries has recently been systematically studied. When contacted by superconducting electrodes, normal (Au) [5] and magnetic (Co and Ni) [46] nanowires acquire superconductivity via the proximity effect. A counter intuitive phenomenon known as the anti-proximity was also observed where the superconductivity of thin Zn and Al nanowires was quenched or weakened when they were contacted by superconducting electrodes [14,23,47]. The effect of normal electrodes on single crystal superconducting nanowires, specifically in enhancing phase slips in the nanowires has not been systematically studied by standard four-probe measurement.

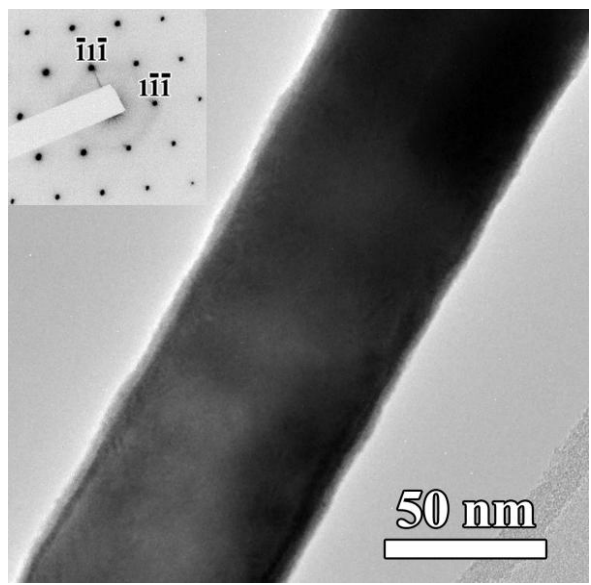

FIG. 1. TEM image of a typical Pb nanowire. The inset shows [110] zone pattern from the same wire.

In this paper, we report such a study of the transport properties of individual single crystal superconducting Pb nanowires of 55 and 70 nm diameter contacted by four normal Pt electrodes. The diameters of these wires are on the order of the coherence lengths of Pb. Interestingly, *R-T* scans, *R-H* measurements and *R-I* curves show a series of resistance steps with increasing temperature, magnetic field, and excitation current respectively as the wires are brought toward the normal state. We attribute these observations to the weakening of superconductivity in nanowires induced by proximity effect of the normal Pt electrodes on the Pb nanowires.

## II. RESULTS and DISCUSSION

The Pb nanowires used in this work were electrodeposited in commercially available track-etched polycarbonate membranes that are coated with Au on one side [13]. The electrolyte $Pb(NH_2SO_3)_2$ was prepared by reacting lead carbonate ($PbCO_3$) with excess sulphamic acid solution in purified $H_2O$ (resistivity>18MΩcm). The transmission electron microscopy (TEM) and selected area electron diffraction study showed that the Pb nanowires are single crystalline (see Fig. 1). It is noticed that there is a 3-4 nm thick oxide shell surrounding the nanowires, which protects the nanowires from getting damaged during the attachment of the electrodes. To measure an individual Pb nanowire, a drop of the nanowire suspension solution is placed on a silicon substrate with a 1 μm thick $Si_3N_4$ insulating layer. The sample is then transferred into a commercial focused ion beam (FIB) etching and deposition system (FIB/SEM FEI Quanta 200 3D) for the attachment of electrodes. As shown in the inset of Fig. 2(a) which is a scanning electron microscopy (SEM) image of the 55 nm diameter sample, four FIB-assisted Pt electrodes are deposited to contact the individual Pb nanowire for a standard four-probe measurement. The distance between the inner edges of the two voltage electrodes of the 55 nm sample is 3.7 μm. In the process of electrode preparation, the deposition current is set to be lower than 10 pA to reduce the destruction and contamination. Transport measurements are carried out in a physical property measurement system (PPMS-Quantum Design).

The results shown in Fig. 2 are obtained from 55 nm and 70 nm diameter nanowires. The distances between two inner edges of the two voltage electrodes of the 70 nm and 55 nm samples are 1.9 μm and 3.7 μm respectively. Figure 2(a) shows the *R-T* curve of an individual 55 nm Pb nanowire measured with excitation current of 50 nA from 1.8 to 300K at zero magnetic field (*H*). The resistance drops of the two wires below the superconducting transition temperature ($T_C$) of Pb, expressed as the resistivity ($\rho$) of the wires, are shown more clearly in Figs. 2(b) and 2(c). The excitation current employed in these measurements is 500 nA. The magnetic field in these curves was aligned perpendicular to the nanowires. The curves in Fig. 2(b) of the 55 nm wire show two obvious resistance drops at 7.0 K and 4.9 K. For the first step between 6.5 K and 7.0 K, the resistance decreases 14% of the normal state value, followed by another more gradual step near 4.9 K. Both steps move to low temperature with increasing field. The wire appears to be normal under a field of 20 kOe at 2K. The $\rho$-*T* curves of 70 nm nanowire (Fig. 2(c)) show three steps at 6.98 K, 5.90 K and 4.67 K. The general trend of the features in these curves is consistent with those of 55 nm sample (Fig. 2(b)). It is reasonable to attribute the resistance drops near 7.0 K found for both wires to the 'intrinsic' superconducting transition of the Pb nanowires because that the $T_C$ of bulk Pb is 7.2 K. What then is the origin of the resistance steps well below $T_C$? TAPS and QPS could induce broadening of the transition and residual resistance near and below $T_C$. However, it is found that when the resistance in normal state ($R_N$) is lower than the quantum resistance ($R_q = h/(2e)^2 \approx 6.5 k\Omega$), the QPS tunneling is strongly suppressed [1]. The $R_N$ of our samples (1.47 kΩ for 55 nm wire and 176 Ω for 70 nm wire) are significantly lower than $R_q$. The TAPS model is expected to be relevant only at temperature very close to $T_C$. The resistance steps in our wires extend to temperatures well below $T_C$ [13]. Thus, intrinsic TAPS and QPS mechanisms are unlikely to be the origin of these steps. The proximity of the wires to the normal Pt electrodes should weaken the superconducting of the Pb nanowires and may account for the residual resistance at temperatures below $T_C$ of Pb. However, it is unexpected and certainly interesting that this proximity effect results in discrete resistance steps. The resistivities of the 70 nm and 55 nm Pb nanowires at room temperature are $26 \times 10^{-6}$ Ω cm and $94 \times 10^{-6}$ Ω cm respectively. These numbers are on the same order as the resistivity of bulk Pb ($21.3 \times 10^{-6}$ Ω cm). The larger $\rho$ of the thinner wire is probably the effect of enhanced surface scattering. We noted that in our four-probe measurement configuration, the contact resistance can be neglected.

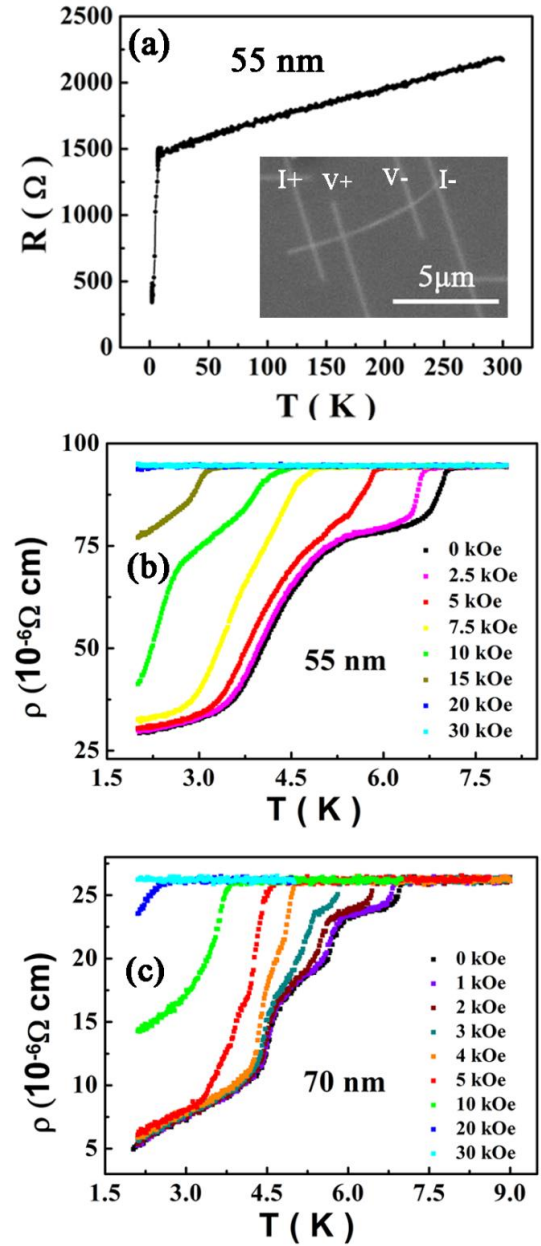

FIG. 2. (Color online) (a) Resistance vs temperature of 55 nm Pb nanowires in the wide temperature range. Inset is the SEM image of the four electrodes measurement; (b) and (c) Resistivity vs temperature of 55 and 70 nm Pb nanowires near and below the $T_C$ in different magnetic fields.

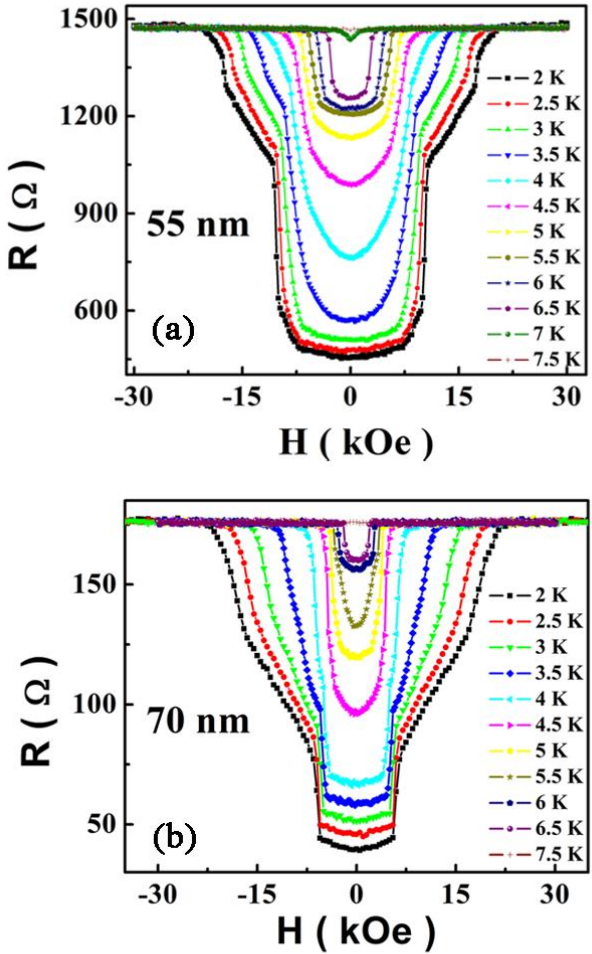

FIG. 3. (Color online) Magnetoresistance *vs* magnetic field of 55 and 70 nm Pb nanowires at different temperatures.

Figure 3 shows the resistance of the Pb nanowires as a function of the magnetic field (*H*) applied perpendicularly to the nanowires at different temperatures. The applied excitation current is 500 nA for 55 nm wire and 1 µA for 70 nm wire. Consistent with the *R-T* scans at different field, the *R-H* scans shown in Figs. 3a and 3b for the 55 and 70 nm wires, display a series of resistance steps with increasing field. The *R-H* scans of the two wires show the same general characteristics. Sharp and well defined resistance steps are found at low temperature. The first step was found near 10 kOe for the 55 nm wire and 7.2 kOe for the 70 nm wire. The resistance steps in the *R-H* scans at different temperatures are consistent with the steps found in the *R-T* scans at different field values. The field at which the two wires are driven into the normal state are almost same (21 kOe) but much larger than that of the bulk Pb (0.803 kOe at zero temperature, 0.74 kOe at 2.0 K). This enhancement in the critical field is a well-studied phenomenon in nanoscale superconductors [48]. With increasing temperature, the critical field decreases and the steps become less well defined and rounded.

The *R-I* curves of the two Pb nanowires measured at different temperatures under zero field are shown in Figs. 4(a) and 4(c), the measurement under different perpendicular magnetic fields at 2 K are shown in Figs. 4(b) and 4(d). The corresponding *V-I* scans under zero field at different temperatures and the 2 K scans under different fields of the 70 nm wire are shown in Fig. 4(e) and Fig. 4(f). These scans show that the increase in resistance and voltage with increasing current is punctuated by sharp steps. Fig. 4(d) shows that the resistance of 70 nm sample at 2 K reaches almost zero at the low current limit of our measurement at 50 nA, but the 55 nm sample (in Fig. 4(b)) shows a residual resistance of about $10^2$ Ω. Unfortunately, we were limited by our equipment from extending the measurement to lower current and temperature. The normal state resistance of 180 Ω of the 70 nm wire at 2K and zero field is reached with stepwise increase in resistance at 50 nA, 5.40 µA, 9.47 µA, 13.00 µA, and 25.47 µA. At higher temperatures, the first step is no longer found and the other steps move to lower current values. Under a field of 2.5 kOe at 2K, the resistance steps also moved to lower current values (Fig. 4(d)). Similar behaviors are found in the *R-I* scans for the 55 nm wire. Similar dependence of these 'critical' current like resistance steps on temperature and magnetic field have been reported in superconducting whiskers [43,44], microbridges [45], and nanowires [13,23, 40-42]. These resistance steps at different current can be understood as a consequence of weak spots along the wire with low local critical currents which act as PSCs. The *R-I* and *V-I* scans in Fig. 4 show that as the excitation current is increased to the critical value of the first "weak spot", a voltage jump (and resistance step) appears in *V-I* curve (*R-I* curve). When the current is increased to the critical value of another 'weak spot', a second step appears. In whiskers, microbridges and nanowires contacted by superconducting electrodes, the PSCs are found at the position of defects [13,45]. In our situation, the proximity of the Pt electrodes is very likely responsible in enhancing the "weak spots" or PSCs in the Pb nanowires. Comparing the *R-I* curves of the 55 nm (3.7 μm long) and 70 nm (1.9 μm long) wires, the steps are much more clearly defined for the shorter 70 nm wire. This is consistent with our proposal that the observed PSCs resistance and voltage steps are induced or enhanced by the normal Pt electrodes since the influence of the proximity effect is expected to be serious in the shorter wire.

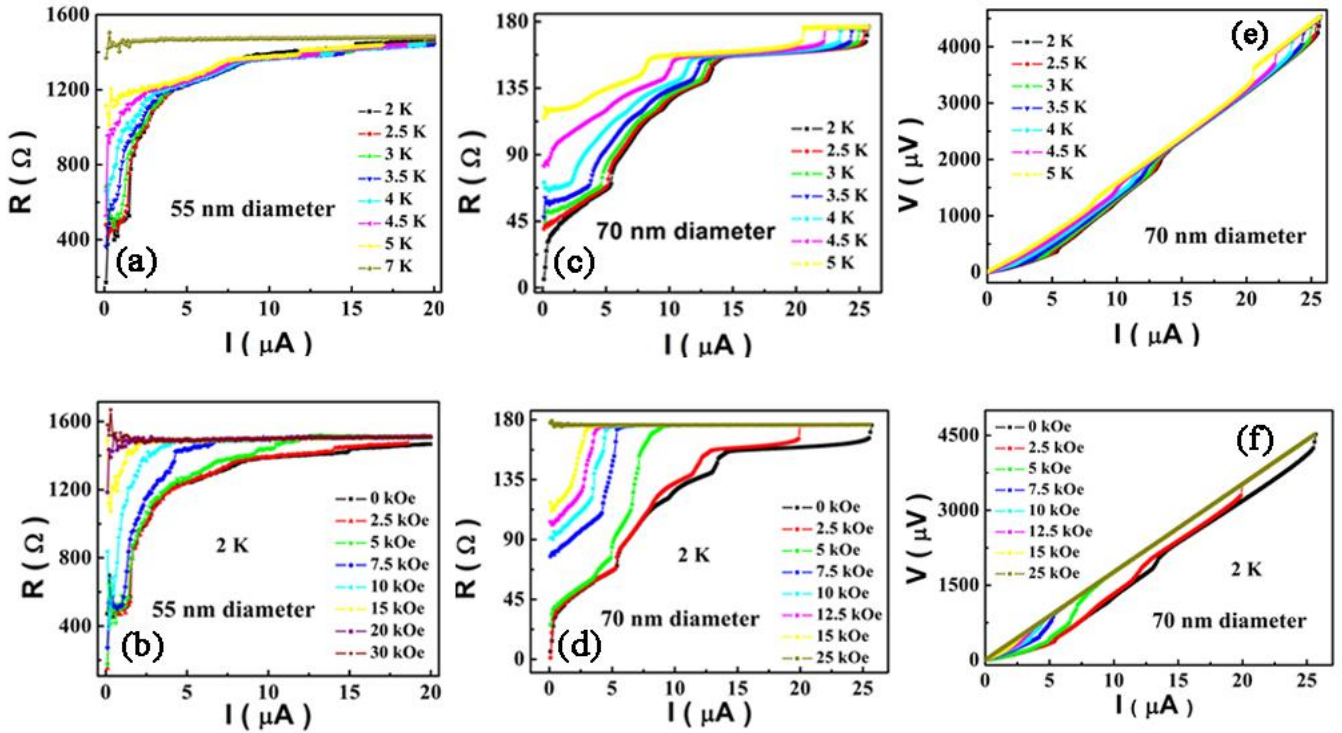

FIG. 4. (Color online) Resistance vs current of 55 and 70 nm Pb nanowires at different temperatures [(a) and (c)] and in different magnetic fields [(b) and (d)]. (e) and (f): voltage vs current curves of 70 nm Pb nanowires at different temperatures and magnetic fields.

### III. CONCLUSIONS

In summary, single crystal Pb nanowires with two different diameters were fabricated by electrochemical deposition. *R-T*, *R-H* and *R-I* curves measured by standard four-probe configuration show a series of resistance steps with increasing temperature, magnetic field, and excitation current respectively as the superconducting nanowires are approaching the normal state. We attribute these steps to PSCs in the Pb nanowires enhanced by the proximity to the normal Pt electrodes.

### ACKNOWLEDGMENT

This work was financially supported by the Penn State MRSEC under NSF grant DMR-0820404, National Basic Research Program (NBRP) of China (No. 2012CB921300), the National Natural Science Foundation of China (No. 11174007), the National Key Basic Research of China under Grant No. 2011CBA00111 and China Postdoctoral Science Foundation (No. 2011M500180).

* jianwangphysics@pku.edu.cn (Wang);
  yisun@pku.edu.cn (Sun);
  chan@phys.psu.edu (Chan).